\documentclass[aps,prb,twocolumn,floatfix,showpacs,superscriptaddress]{revtex4}

\usepackage{graphicx}
\usepackage{bbm}
\usepackage{amsmath,amssymb}

\newcommand{\be}{\begin{equation}}
\newcommand{\ee}{\end{equation}}
\newcommand{\bea}{\begin{eqnarray}}
\newcommand{\eea}{\end{eqnarray}}
\renewcommand{\vr} {{\bf r}}

\begin{document}
%\title{Self-consistent energies and densities for flat electronic systems} 
\title{Large two-dimensional electronic systems: \\ Self-consistent energies and densities at low cost}
\author{E. R{\"a}s{\"a}nen}
\email[Electronic address:\;]{esa.rasanen@tut.fi}
\affiliation{Department of Physics, Tampere University of Technology, FI-33101 Tampere, Finland}
\affiliation{Nanoscience Center, Department of Physics, University of
  Jyv\"askyl\"a, FI-40014 Jyv\"askyl\"a, Finland}
\affiliation{Physics Department, Harvard University, Cambridge, Massachusetts 02138, USA}
   \author{S. Pittalis}
\affiliation{Department of Physics and Astronomy, University of California, Irvine, California 92697, USA}
\affiliation{Department of Physics and Astronomy, University of Missouri, Columbia, Missouri 65211, USA}
\author{G. Bek\c{c}io\u{g}lu}
\affiliation{Nanoscience Center, Department of Physics, University of
  Jyv\"askyl\"a, FI-40014 Jyv\"askyl\"a, Finland}
\affiliation{Dahlem Center for Complex Quantum Systems, Physics Department, Freie Universit\"at Berlin, Arnimallee 14, 14195 Berlin, Germany} 
    \author{I. Makkonen}
\affiliation{COMP Centre of Excellence, Helsinki Institute of Physics and 
Department of Applied Physics, Aalto University 
School of Science, P.O. Box 14100, FI-00076 AALTO, Espoo, Finland} 
%   \author{K. Burke}
%\affiliation{Department of Physics and Astronomy, University of California, Irvine, California 92697, USA} 
\date{\today}

\begin{abstract} 
We derive a self-consistent local variant of the Thomas-Fermi approximation for 
(quasi-)two-dimensional (2D) systems by localizing the Hartree term. The scheme results 
in an explicit orbital-free representation of the electron density and energy in terms of the 
external potential, the number of electrons, and the chemical potential 
determined upon normalization. We test the method over a variety 
2D nanostructures by comparing to the Kohn-Sham  2D-LDA calculations up to 
600 electrons. Accurate results are obtained in view of the negligible computational 
cost. We also assess a local upper bound for the Hartree energy.

\end{abstract}

 
\maketitle

\section{Introduction}
Orbital-free density-functional theory (OF-DFT) is a 
computationally appealing method to deal with large 
systems beyond the reach of conventional DFT. 
At present, OF-DFT methods can handle systems 
up to a million atoms.~\cite{HungCarter} 
These methods await to be fully explored in the context 
of low-dimensional systems and nanoelectronic devices. 
In two-dimensional (2D) physics one of the main challenges
of DFT is to deal with regions of the 2D electron 
gas~\cite{vignale} comprising hundreds or thousands of
interacting electrons, e.g., in the quantum Hall regime.~\cite{qh}
%%%%%%%%%%%%%%%%%%%%%%%%%%%%%%%%%%%%%%%%%%%%%%%%%%%%%%%%%%%%%%%%%%%%%%%%%%%%%%%%%
% STEFANO: I remove the stuff below because it sounds like we need Exc... but then later we kind of trash it all by localizing TF... I moved up \cite{qh} just above %%
%%%%%%%%%%%%%%%%%%%%%%%%%%%%%%%%%%%%%%%%%%%%%%%%%%%%%%%%%%%%%%%%%%%%%%%%%%%%%%%%%
%During the recent years successful efforts have been made
%to describe quantum Hall devices by incorporating the
%electron-electron (e-e) interactions up to the TF 
%level,~\cite{qh} but improvements regarding, e.g.,
%the exchange-correlation effects neglected in the TF
%approach are needed.

As the name suggests, OF-DFT
is free from the use of the Kohn-Sham orbitals 
needed in the calculation of the Kohn-Sham kinetic energy,
and thus the only explicitly needed variable is the 
electron density $\rho(\vr)$. The earliest OF-DFT 
method dates back to the Thomas-Fermi (TF) theory employing the 
exact result of the homogeneous electron gas for the 
kinetic energy, and the Hartree approximation for the 
e-e interaction. In fact, most orbital-free 
schemes can be regarded as modifications or
improvements to the TF method.~\cite{of-reviews} 

The crux of the problem in OF-DFT is to find an approximation for the
(non-interacting) kinetic-energy functional that may be generally applicable.
For this ambitious goal, a promising approach is an orbital-free 
formulation exploiting the potential rather than the the 
density as the basic variable.~\cite{kburke1} 
%%%%%%%%%%%%%%%%%%%%%%%%%%%%%%%%%%%%%%%%%%%%%%%%%%%%%%%%%%%%%%%%%%%%%%%%%%%%%%%%%
% STEFANO: the added part just below is important because it allows us to cite people that understand the topic and to make the reader fully comfortable about the kinetic part %%
%%%%%%%%%%%%%%%%%%%%%%%%%%%%%%%%%%%%%%%%%%%%%%%%%%%%%%%%%%%%%%%%%%%%%%%%%%%%%%%%%
In 2D the TF approximation for the kinetic energy works remarkably well -- 
in fact, the gradient corrections to it vanish to 
{\em all} orders.~\cite{old19,old20,old21,antti,old22}
Moreover, for the 2D Fermi gas in harmonic trap the TF kinetic energy 
yields the exact noninteracting kinetic energy when the 
exact density is used as the input.~\cite{old19} In this work, we
replace the Hartree term with a much simpler {\em local} expression 
that significantly speeds up the calculations.

Recently, a 2D orbital-free expression for the energy was shown 
to lead to a major improvement over the TF results when applied
to quantum dots and slabs up to 200 interacting electrons.~\cite{old}
The energies were not obtained self-consistently, but 
by using the electron densities from the 2D local-density
approximation (2D-LDA). 
The expression of the functional is
\be
E_{\rm tot}[\rho(\vr)] = T_{\rm TF}[\rho(\vr)]+W[\rho(\vr)]  + \int d \vr\, \rho(\vr)\,v_{\rm  ext}(\vr),
\label{etot}
\ee
where
\be
T_{\rm TF}[\rho(\vr)] = \frac{\pi}{2} \int d \vr\, \rho^2(\vr)
\label{tf}
\ee 
is the TF kinetic energy in 2D and the last term is the energy 
contribution due to the external scalar (confining) potential 
$v_{\rm  ext}(\vr)$. The total electron-electron interaction energy 
is given by
\be
W[\rho(\vr)] = \frac{\pi}{2} \sqrt{\frac{N-1}{2}} \int d \vr \,\rho^{3/2}(\vr)\;,
\label{w}
\ee
which was obtained from a crude approximation by using a Gaussian
ansatz for the {\em cylindrical} average of the
pair density, and enforcing an overall linear behavior under 
uniform coordinate scaling.
This was also partially inspired by the fact that an analogous 
Gaussian anasatz for the one-body reduced-density matrix 
eventually leads to highly accurate exchange 
energies.~\cite{gaussianapproxpaper}

In Eqs.~(\ref{etot}) and (\ref{w}) it is apparent that for $N=1$ the 
functional reduces to the {\em noninteracting} TF approximation.
For $N \ne 1$ the interaction contribution 
is similar to the form of the exchange energy in the 2D-LDA, 
but with a different prefactor $\sqrt{N-1}$. 
This approximation completely eliminates the 
computational burden of the Hartree term in the 
TF approximation.
We point out that Eq.~(\ref{etot}) has been employed by 
others to compute the total energy of a realistic semiconductor 
quantum dot formed in gate- and etching-defined devices.~\cite{Siddiki2011}
The results confirm the good balance between accuracy and efficiency
of the functional.

In this work, we verify that the same approach works well 
in a fully self-consistent framework. In addition,
an important modification improves its performance.
As remarked in Ref.~\onlinecite{old}, 
the derivation of Eq.~(\ref{w}) employs,
among others, a Hartree-Fock expression
for the pair density that leads to a
particular choice for the overall 
coefficient. Here we propose a different 
coefficient that has a non-empirical justification
as explained below. Our proposal also leads to
an estimate for a {\em local} upper bound of the Hartree
energy in 2D systems. Finally we test the derived approximation 
self-consistently for an extensive set of 2D systems.
Remarkably accurate results for the total energy density are 
obtained in view of the simplicity of the scheme
and the negligible computational cost.

%The paper is organized as follows:

\section{Theory}

\subsection{Upper bound for the Hartree term}\label{hartree}

Lieb and co-workers have proved that the TF theory 
is asymptotically exact for large quantum dots.~\cite{lieb} 
In addition, Burke and co-workers have shown that successful 
DFT approximations are those that become {\em asymptotically} 
exact for the class of systems under investigation.~\cite{kburke2} 
It is natural to follow the same principle in 2D, where
semiconductor quantum dots play the role of ``artificial atoms''.
The fundamental question is whether the form in Eq.~(\ref{etot}) can be made, 
in some sense, asymptotically correct. We provide an affirmative 
answer by exploiting the existence of a {\em local} upper bound for the 
Hartree energy:
\be\label{upperb}
\frac{	1}{2} \int d \vr \int d \vr' \frac{\rho(\vr)\rho(\vr')}{|\vr - \vr'|} \le \frac{	1}{2} C \sqrt{N} \int d \vr \rho^{3/2}( \vr)\;,
\ee
where $C$ is a constant to be estimated.~\cite{Pino} 
The latter expression suggests that, for large $N$,  we may be able 
to energetically approach an exact (TF) theory ``from above''.
Of course, sole energy bounds do not allow us to directly control
the behavior of the functional derivatives of the obtained approximations.
This is expected to affect the accuracy of self-consistent densities.
Moreover, the considered bound does not allow size-consistence,
which may have severe effects on the chemical potentials as well
as on ``multi-center'' systems with separated confining potentials.

In order to find an ``optimal''  constant $C$ in
Eq.~(\ref{upperb}), we consider a harmonically confined quantum dot,
where the confinement potential is kept fixed while
adding more electrons. Eventually, the density becomes  
relatively flat, resembling a disk with radius $R$.
In the large-$N$ regime the Hartree energy dominates over
the exchange and correlation energy. For example,
for $N=6$, 30, and 600 we have $|E_{xc}|/E_H \approx$ 0.33,
0.15, and 0.03, respectively, with the oscillator strength
$\omega=0.5$.
For a flat 2D density, the Hartree energy is given exactly by~\cite{seidl}
\be\label{Hdisk}
E^{\rm disk}_H = \frac{8}{3\pi} \frac{N^2}{R}.
\ee
Combining Eqs.~(\ref{upperb}) and (\ref{Hdisk}), 
we see that asymptotically, Eq.~(\ref{upperb}) tends to an 
{\em equality} with $C^{\rm disk}=  \frac{16}{3 \sqrt{\pi}}$.
Previously, taking as reference Gaussian-like density, 
it was suggested that $C = \frac{3\pi}{\sqrt{2}}$, i.e., a value 
significantly greater than $C^{\rm disk}$. Therefore, we could
expect that our result may work as a good estimation for the 
upper bound of $E_H$.

\begin{figure}
\includegraphics[width=0.9\columnwidth]{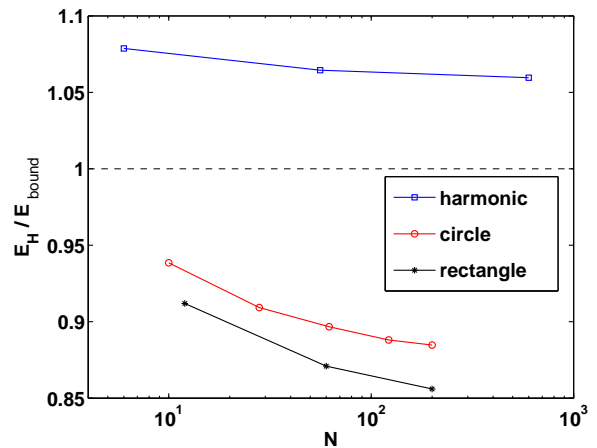}
\caption{(color online) 
Hartree energy $E_H$ from a self-consistent 2D-LDA calculation 
with respect to our estimate for the upper bound $E_{\rm bound}$, 
i.e., the r.h.s. of
Eq.~(\ref{upperb}) with $C=\frac{16}{3 \sqrt{\pi}}$. Results
as a function of $N$ are shown for three different 2D systems.
}
\label{bounds}
\end{figure}

Figure~\ref{bounds} shows the Hartree 
energy $E_H$ from a self-consistent 2D-LDA calculation 
with respect to our upper bound estimate $E_{\rm bound}$. 
Three different quantum-dot systems are considered: 
(i) harmonic dots defined by $v_{\rm ext}(r)=\omega^2 r^2/2$
with $\omega=0.5$, (ii) circular dots defined by a hard-wall
potential [$v_{\rm ext}(r)=0$ at $r\leq R$ with $R=10$, 
$v_{\rm ext}(r)\rightarrow\infty$ at $r>R$], and (iii)
rectangular hard-wall dots~\cite{recta1,recta2} with side lengths $L$ and $2L$,
where $L=10$. We find that in all the cases, and with different
$N$, the Hartree energy is very close to our suggested bound. However, 
it is noteworthy that our estimate does not serve as 
a tighter upper bound (rather as an approximation for it), 
as for for harmonic dots we obtain
values above $E_H$. 

\subsection{Self-consistent scheme}\label{sc}

First, for  large $N$ we have $\sqrt{N-1} \approx \sqrt{N}$, and 
therefore -- and according to the above analysis -- 
Eq.~(\ref{etot}) can be modified as follows:
\be
E^{\alpha}_{\rm tot}[\rho(\vr)] = T_{\rm TF}[\rho(\vr)]+ \alpha W[\rho(\vr)]  + \int d \vr\, \rho(\vr)\,v_{\rm  ext}(\vr)
\label{NEWetot}
\ee
where $\alpha = \frac{8}{3}(\frac{2}{\pi})^{3/2} \approx 1.35453$.
It is apparent that the modified form fails to to recover the 
size-consistency of the TF approximation.
An immediate consequence (that can be 
verified analytically for, e.g., rectangular quantum dots) 
is that the fundamental 
relation $\mu = d E_{\rm tot}/d N$ is not fulfilled.
However, in our numerical tests below we have not found severe 
consequences of this deficiency.

%Quantum slabs as defined above are also useful model systems 
%in the analytic and self-consistent application of our
%approximation in limiting situations. In contrast
%with the analysis above (Fig.~\ref{fig1}) we now 
%include the e-e interactions. First, when  the area $A$ of 
%the slab is fixed and $N$ goes to infinity we get
%$\mu\approx\left[\pi/A+3/(4\sqrt{2})\right]N$ and 
%$E_{\rm tot}\approx (\mu/2+N/8)N$. 
%Then, when $N >> 1$ but finite and $A$ goes to infinity, we get
%$\mu\approx 3\pi N/(4\sqrt{2A})$ and $E_{\rm tot}\approx 2\mu N/3$.
%Finally, when both $N$ and $A$ approach infinity while  
%keeping the (self-consistent) density constant, we get,
%again, $\mu\approx 3\pi N/(4\sqrt{2A})$ and 
%$E_{\rm tot}\approx 2\mu N/3$.
%In all these three limiting cases, our approximation 
%apparently {\em fails} to satisfy the fundamental relation
% We attribute this fact 
%to the fact that $W$ is not size-consistent.~\cite{old} 

In order to find the ground-state density we have to minimize 
Eq.~(\ref{NEWetot}) for a fixed number of particles. 
We may first
express the total energy in a single integral as
\bea
E^\alpha_{\rm tot}[\rho(\vr)] & = & \int d \vr\, F[\rho(\vr)] = \int d \vr\,\Bigg[\frac{\pi}{2} \rho^2(\vr) \nonumber \\ 
& + & \frac{\pi\alpha}{2}\sqrt{\frac{N-1}{2}}\,\rho^{3/2}(\vr)+ \rho(\vr)\,v_{\rm  ext}(\vr)\Bigg].
\label{etot2}
\eea
We have to find a stationary value for the functional 
$F[\rho(\vr)]$ with respect to variations in $\rho(\vr)$.
To take the electron number conservation into account, we introduce 
another functional
$G[\rho(\vr)]=\rho(\vr)$ so that
\be
\int d \vr\, G[\rho(\vr)] = N.
\label{norm}
\ee
This constraint introduces a Lagrange multiplier $\mu$ in the 
variational equation, which can be written as
\be
\frac{d F}{d \rho}-\mu\frac{d G}{d \rho} = 0.
\ee
Substituting $F$ and $G$ to this equation yields
\be
\pi\rho(\vr)+\frac{3\pi\alpha}{4}\sqrt{\frac{N-1}{2}} \rho^{1/2}(\vr)+v_{\rm ext}(\vr)-\mu = 0.
\label{den}
\ee
As this expression is quadratic in $\rho^{1/2}$, we find an explicit
expression for the density,
\bea
\rho(\vr) & = & \Bigg\{-\frac{3\alpha}{8}\sqrt{\frac{N-1}{2}} \nonumber \\
& + & \frac{1}{2}\sqrt{ \left[ \frac{9\alpha^2}{32}(N-1)-\frac{4}{\pi}\left[v_{\rm ext}(\vr)-\mu\right] \right]_{+}}\Bigg\}^2.
\label{density}
\eea
This shows that 
the density can be solved instantaneously for any external 
potential $v_{\rm ext}$ and any $N$. The only variable to be 
determined numerically is $\mu$ that follows from the normalization 
condition in Eq.~(\ref{norm}).
The symbol $\left[ ... \right]_{+}$ in Eq.~(\ref{density})
represents an additional constraint that no sign 
changes under the square in Eq.~(\ref{density}) (leading to 
unphysical ``nodal lines'' in the density), nor negative values 
under the square-root (leading to complex densities), are allowed. 
Once $\rho(\vr)$ is determined from Eq.~(\ref{density}), 
the total energy is finally obtained from Eq.~(\ref{NEWetot}).

{Let us emphasize the difference between the present and
and the TF approximation in a practical sense}. In the latter, the variational procedure
applied to the total energy leads to an {\em integral} equation 
for the density. The TF scheme then transforms into a differential 
equation (which in 3D leads to the Poisson equation). 
Instead, our functional is free from this complexity due to the simple
expression for the interaction energy [Eq.~(\ref{w})] in comparison 
with the Hartree integral utilized by the TF method. Although the
Hartree term is simple to calculate in most applications, it may
become a bottleneck in large 2D systems. In any case, 
it is appealing to have a method for the first approximation
of the electronic density and energy with a negligible
computational cost.

\section{Applications and analysis}

Next we test the self-consistent scheme of the previous section
for a set of 2D quantum dots
and rings including the e-e interactions. 
We use DFT with the 2D-LDA~\cite{attaccalite} as 
our reference method; in the range of systems and parameters 
considered here the LDA has been shown to provide -- for the
present purpose -- sufficiently accurate
total energies and densities (see, e.g., Refs.~\onlinecite{saarikoski}
and \onlinecite{bigring} for quantum dots and rings, respectively).
The LDA calculations are performed using the 
{\tt octopus} code~\cite{octopus} as well as another
code exploiting circular symmetry.~\cite{iljapaper}

\begin{figure}
\includegraphics[width=0.9\columnwidth]{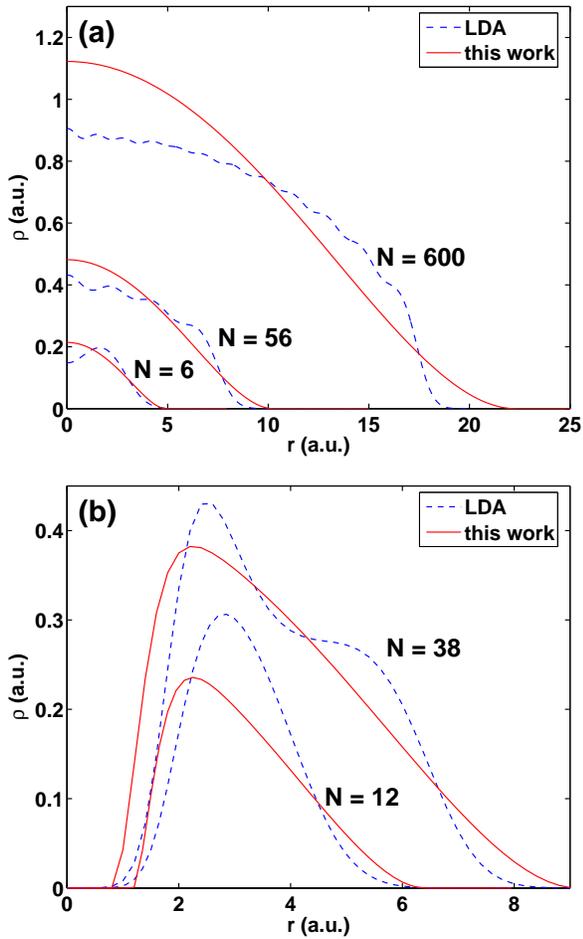}
\caption{(color online) (a) Electron densities in harmonic
quantum dots containing $N=6$, 56, and 600 electrons,
respectively. The dashed lines denote the
DFT results within the 
2D-LDA. The solid lines correspond to the results of the present
orbital-free functional. (b) The same as in (a) but for
two quantum rings containing $N=12$ and 38 electrons,
respectively.
}
\label{fig2}
\end{figure}

Figure~\ref{fig2}(a) shows the radial densities obtained 
from Eq.~(\ref{density}) for 2D
harmonic quantum dots with $\omega=0.5$ 
and $N=6$, 56, and 600 (solid lines). The dashed lines show
the corresponding LDA results. As expected, the present functional
cannot describe the shell structure due to the lack of
orbitals. However, the overall shape of the density profile
is satisfactory in a qualitative fashion, and the 
correct radial extent of the density profile is obtained in 
all cases. 
%Note that the correct 
%density has a Gaussian decay,~\cite{bj} whereas in our
%approximation, by definition, the decay follows the external 
%potential up to the power of one. 
%As an obvious 
%drawback, agreement with the LDA does not seem to improve as a 
%function of $N$.

In Fig.~\ref{fig2}(b) we show the corresponding result for a
quantum ring modeled by $v_{\rm ext}(r)=\omega^2 r^2/2 + V_0 \exp(-r^2/d^2)$
with $\omega=0.5$, $V_0=20$, and $d=1$. The model potential 
is the same as the one used in Refs.~\onlinecite{bigring}, 
\onlinecite{ring}, and \onlinecite{newring},
the last reference showing direct comparison with an experiment. 
We find a reasonable qualitative agreement between the present 
functional and the LDA. The qualitative agreement is
similar for both $N=12$ and $N=38$.

\begin{figure}
\includegraphics[width=0.99\columnwidth]{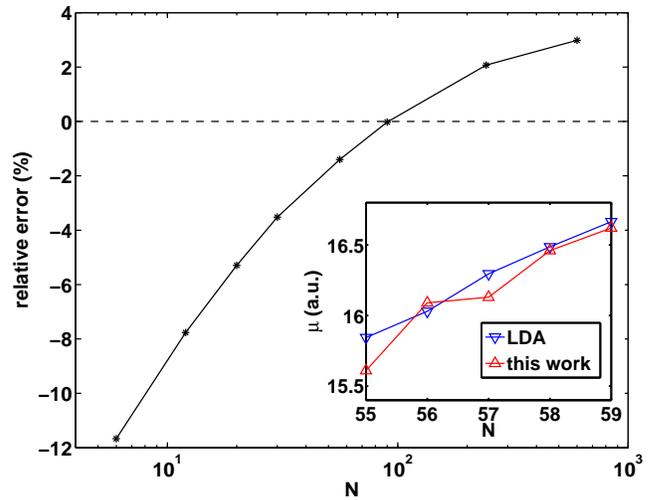}
\caption{(color online) Relative error in the total energies
of harmonic quantum dots
calculated with our self-consistent scheme with respect
to 2D-LDA results. The inset shows the obtained
chemical potentials $\mu(N)=E_{\rm tot}(N)-E_{\rm tot}(N-1)$ for $N=55\ldots 59$
in comparison with the 2D-LDA.
}
\label{para}
\end{figure}

Apart from densities, it is important to assess the
performance of the present functional for total energies.
Figure~\ref{para} shows the relative total-energy 
differences from the reference 2D-LDA results for a set
of harmonic quantum dots up to $N=600$. Overall, the
accuracy is remarkably good in view of the negligible
computational cost. Even for small $N$ the accuracy is
well handled, e.g., with $N=12$ the relative error
is below $8\%$, which is considerable improvement over
the TF approximation.~\cite{old} However, the main
interest for practical applications is in the large-$N$
regime. For $N=600$ our approximation overestimates the
total energy only by $\sim 3\%$. According to Fig.~\ref{para}
the error then increases with $N$, but most likely saturates.
It can be extrapolated that for $N\sim 10000$ the error 
of our approximation is still under $5\%$ (note the log-scale
in the x-axis).

The inset of Fig.~\ref{para} shows the obtained (spin-compensated)
chemical potentials $\mu(N)=E_{\rm tot}(N)-E_{\rm tot}(N-1)$ for 
$N=55\ldots 59$ in harmonic quantum dots. In view of
the lack of size-consistency (see Sec.~\ref{sc}), our scheme yields
surprisingly accurate results in comparison with the 2D-LDA.
We point out, however, that due to the lack of orbital dependency
(and thus the shell structure) we can only obtain the qualitative
behavior of $\mu$ without any detailed features.

\begin{figure}
\includegraphics[width=0.8\columnwidth]{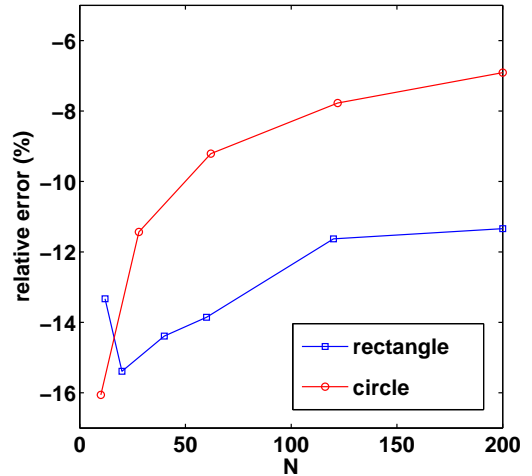}
\caption{(color online) Relative error in the total energies
for circular and rectangular hard-wall quantum dots 
calculated with our self-consistent scheme with respect
to 2D-LDA results.
}
\label{hard}
\end{figure}

In Fig.~\ref{hard} we compare our results for hard-wall
circular and rectangular quantum dots with 2D-LDA results. The
parameter values are the same as in Sec.~\ref{hartree}.
Overall, the errors are slightly larger than for harmonic
quantum dots. On the other hand, the errors become smaller
with $N$ so that we can expect reliable results at least
within $N\sim 10^3\ldots 10^4$. We point out that real-space 
2D-LDA calculations are numerically tedious in those systems.
Detailed assessment of our scheme in the very large-$N$
limit ought to be performed with respect to TF results.

Finally we discuss the relevance of $\alpha$ in terms of
the performance of our scheme. As described in Sec.~\ref{hartree},
the inclusion of $\alpha$ results from the limit of
a circular and flat 2D density whose Hartree energy is known.
Eventually, $\alpha$ appears in Eq.~(\ref{NEWetot}) as a prefactor
of $W$. Instead, in a previous non-self-consistent
formulation~\cite{old} $\alpha$ was equal to one on the basis
of the Hartree-Fock expression for the pair density.
For completeness, we have tested
our self-consistent scheme also with $\alpha=1$ and found
drastically worse results. For example, for a harmonic quantum 
dot with $N=600$ and for a circular hard-wall dot with $N=200$ 
the formulation with $\alpha=1$ yields -- in both cases -- 
a $20\%$ overestimation of the total energy. 
In contrast, and as shown above, the present approach
(with $\alpha = \frac{8}{3}(\frac{2}{\pi})^{3/2} \approx 1.35453$)
yields respective errors of $3.0\%$ and $6.9\%$.
Therefore, the inclusion of $\alpha$ can be also practically 
validated.

\section{Conclusions}

We have derived a self-consistent scheme
to compute approximate electron densities and total 
energies for confined (quasi-)two-dimensional (2D) systems. 
Our scheme can be applied to any
number of electrons with a negligible computational
cost.  In view of its extreme simplicity, we
have obtained appealing results for electron
densities and total energies in a variety of systems (such as harmonic
and hard-wall quantum dots and quantum rings).
Preceding the derivation of our self-consistent
scheme, we have found a good approximation for an 
upper bound of the Hartree energy in 2D.
The present scheme may be useful in negligble-cost computational 
investigations of 2D systems such as quantum Hall devices.

\begin{acknowledgments}
This work was supported by the Academy of Finland
(E.R. and G.B.), Wihuri Foundation (E.R.), ERASMUS
Internship Programme (G.B.), the DOE grant 
DEFG02-05ER46203 (S.P.), the NSF grant CHE-1112442 (S.P.) and the 
European Community’s FP7 through the CRONOS project, grant 
agreement no. 280879 (E.R.). CSC Scientific Computing Ltd. is acknowledged
for computational resources.
\end{acknowledgments}

\end{document}